\documentclass[12pt,letterpaper]{article}
\usepackage{osajnl}
\usepackage{overcite,hyperref}

\begin{document}

\title{Material effects in airguiding photonic bandgap fibers}
\author{Jesper L{\ae}gsgaard$^1$ Niels Asger Mortensen$^2$ Jesper Riishede$^1$ and Anders Bjarklev$^1$}
\affiliation{$^1$Research Center COM, Technical Univ. of Denmark Bldg. 345v,
 DK-2800 Kgs. Lyngby, Denmark\hspace{8pt}
$^2$Crystal Fibre A/S, Blokken 84, DK-3460 Birker{\o}d, Denmark}
%
%
\begin{abstract}

  The waveguiding properties of two silica-based airguiding photonic bandgap fiber designs
are investigated with special emphasis on material effects. 
The nonlinear coefficients are found to be 1-2 orders of magnitude smaller than
those obtained in index-guiding microstructured fibers with large mode areas.
The material dispersion of silica makes a significant contribution to the
total chromatic dispersion although less than 10\% of the field
energy is located in the silica regions of the fibers. These findings suggest
that dispersion engineering through the choice of base material 
may be a possibility in this type of fibers. 
\end{abstract}
\maketitle
%
%
\newcommand{\be}{\begin{equation}}
\newcommand{\ee}{\end{equation}}
\newcommand{\bea}{\begin{eqnarray}}
\newcommand{\eea}{\end{eqnarray}}
\newcommand{\eps}{\varepsilon}
\newcommand{\om}{\omega}
\newcommand{\dloge}{\frac{d\ln\eps}{d\om}}
\newcommand{\dtloge}{\frac{d^2\ln\eps}{d\om^2}}
\newcommand{\dedw}{\frac{\partial E_d}{\partial\om}}
\newcommand{\dpedw}{\frac{\partial E_d}{\partial\om}}
\newcommand{\dplogeddw}{\frac{\partial \ln E_d}{\partial\om}}
\newcommand{\dbl}{$d/\Lambda$}
\newcommand{\lbl}{$\lambda/\Lambda$}
\newcommand{\aeff}{$A_{\rm eff}$ }

\section{Introduction}
 
Photonic bandgap (PBG) fibers guiding light in a hollow core surrounded by
a cladding structure with a bandgap at a refractive index below the light line
have attracted considerable attention since their first experimental 
demonstration by Cregan et al\cite{cregan1999}. Such fibers have been
proposed as candidates for highly linear and possibly low-loss transmission 
fibers\cite{cregan1999}, devices for particle transport\cite{benabid2002a},
dispersion compensation\cite{ouyang2002} and gas nonlinearity experiments\cite{benabid2002b}. In contrast to conventional fibers, the 
useful wavelength range is not limited by the absorption loss and nonlinearity 
of the base material. The recent fabrication of silica-based airguiding PBG 
fibers with attenuation coefficients below 30 dB/km over a considerable 
wavelength range\cite{venkataraman2002} opens up for a wide range of practical applications. 

  It is usually assumed that the influence of the base material on elementary 
fiber properties are negligible for this class of fibers, however the 
theoretical investigations performed up to now\cite{broeng-ag,white-ag-mp} have primarily focused on 
establishing the shape and transmission windows of the guided modes, and 
have not, except in the case of circular Bragg fibers\cite{ouyang2002},
provided a detailed modeling of key quantities such as group 
velocity dispersion (GVD), nonlinear coefficients etc. The purpose of the 
present work is to model two instances of a simple and well established design of airguiding PBG 
fibers with particular emphasis on the interaction between light and base 
material, which is here assumed to be silica. Specifically, we derive the 
fraction of the field energy present in the silica regions of the fiber, the 
nonlinearity coefficient (expressed as an effective area) arising from 
the material nonlinearity of silica, and the GVD including material
dispersion effects. We demonstrate that the fraction of the field energy
present in the silica is below 10\% for both structures studied, and that
the nonlinearity coefficients arising from silica are 1-2 orders of magnitude
lower than what is obtained in the best silica-based large mode-area fibers.
Furthermore, it will be shown that the GVD is considerably influenced by the 
dispersion of the base material, and demonstrate the reason for this 
interesting effect by a detailed analysis of the material contributions to
the GVD. 

  The rest of the paper is organized as follows: In section 2, we describe the
fiber designs to be investigated and briefly
outline the theoretical approach adopted here, including the basic formulae
for group velocity and nonlinearity coefficients. In section 3, our numerical
results are presented and discussed while section 4 summarizes our conclusions.
 
\section{Theoretical approach}
 
\begin{figure}[h]
\resizebox{13cm}{!}{\includegraphics[0cm,0cm][20cm,29cm]{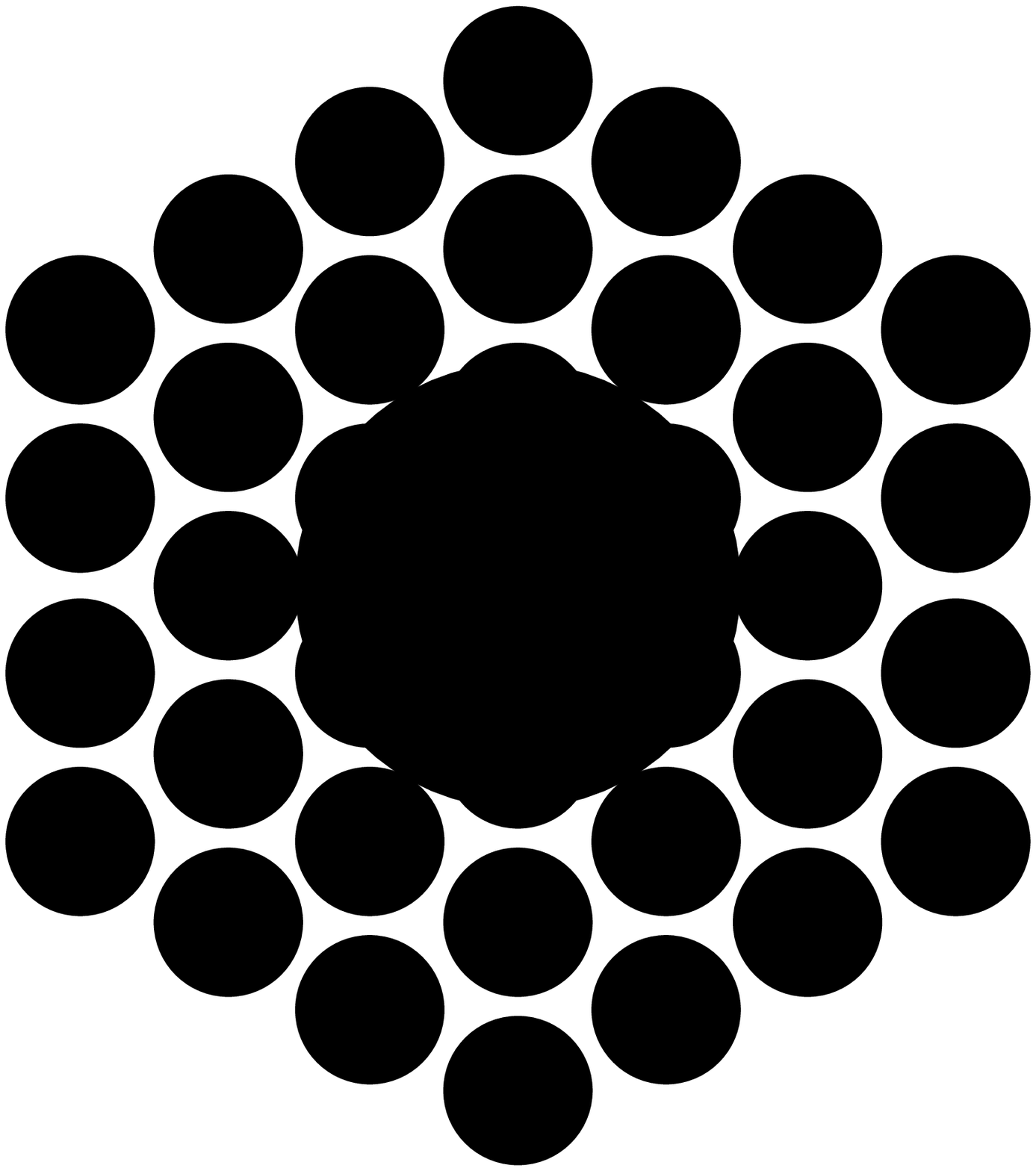}}
\caption{}
\label{figstruct}
\end{figure}

 The two fiber structures to be investigated are both based on a cladding
structure consisting of a triangular lattice of airholes, with a core defined
by a larger airhole. The structure is characterized by three parameters:
The distance between cladding hole centers, $\Lambda$, which is commonly 
denoted the pitch, and the diameters of core and cladding holes. We fix the
core hole diameter at 3$d$, where $d$ is the cladding hole diameter, and
investigate two designs with $d/\Lambda$=0.88 and $d/\Lambda$=0.95 respectively.
A design similar to the former has earlier been modeled by Broeng and co-workers\cite{broeng-ag} whereas
the latter resembles a low-loss airguiding fiber recently fabricated by
Venkataraman and co-workers\cite{venkataraman2002}. A schematic picture of the 
core and nearest cladding region in the design with $d/\Lambda$=0.88
is shown in Fig.~\ref{figstruct}.

  In the present work we solve Maxwells equations by expanding the dielectric
function and magnetic field vector in plane waves using a freely 
available software package\cite{mpb}. Having obtained the magnetic field 
vector the electric fields are straightforwardly calculated by use of Amperes 
law. The adoption of a planewave basis
necessiates the use of periodic boundary conditions, however the interaction
between nearest-neighbor repeated images of the guiding defect can be 
minimized by a proper choice of the transverse Bloch wave vector\cite{matdisp}.
We use a supercell consisting of 8$\times$8 elementary cells of the triangular
lattice comprising the cladding. The Fourier grid used for the plane-wave
expansion has 64$\times$64 meshpoints in each elementary cell for the 
structure with \dbl=0.88, and 96$\times$96 meshpoints for the structure with
\dbl=0.95. With these parameters, the dispersion coefficients and all other
results, are converged within a few percent.

  The nonlinear coefficient of a fiber expresses the change in effective
index of the guided mode arising from nonlinear effects for a given input 
power. The dependency of the nonlinear coefficient on the form of the
guided mode is usually expressed by an effective area\cite{agrawal}:

\vspace{12pt}
\be
\Delta n_{\rm eff} = P\frac{n_2^P}{A_{\rm eff}} \label{aeffdef}
\ee
\vspace{12pt}

Here $P$ is the power launched into the fiber and $n_2^P$ is a material
nonlinear coefficient (related to the third-order nonlinear susceptibility)
in units of W/m$^2$. For conventional, all-silica, fibers \aeff may
be expressed as:\cite{agrawal}

\vspace{12pt}
\be
A_{\rm eff}=\frac{\left(\int \mid {\bf E}\mid^2 dA\right)^2}{\int \mid {\bf E}\mid^4 dA}\label{oldaeff}
\ee
\vspace{12pt}

We have recently shown that for situations in which a substantial part of the
field propagates in air the above definition must be generalized to\cite{pbg_ea}:

\vspace{12pt}
\be
A_{\rm eff} = \left(\frac{n_1}{n_g^0}\right)^2
\frac{\left(\int{\bf E\cdot D} dA\right)^2}{\int_{\rm SiO_2} \mid {\bf E\cdot D}\mid^2dA}  \label{aeff}
\ee
\vspace{12pt}

Note that the integration in the denominator is now restricted to the silica
parts of the fiber.
This formula has been derived without making assumptions about the field energy
distribution and is therefore applicable even in the extreme case of airguiding
PBG fibers. Of course, the \aeff values obtained for these fibers have little
to do with the physical extent of the guided modes, however the expression of
the nonlinear coefficient in this form facilitates the comparison with 
more conventional fibers guiding light in silica or other materials. 

  The GVD coefficient, $D$, is defined as:

\vspace{12pt}
\be
D=-\frac{\om^2}{2\pi c}\frac{d^2\beta}{d\om^2}=
\frac{\om^2}{2\pi cv_g^2}\frac{dv_g}{d\om}
\label{deq}
\ee
\vspace{12pt}

where $v_g$ is the group velocity:

\vspace{12pt}
\be
v_g=\frac{d\om}{d\beta}
\ee
\vspace{12pt}

  In the present case, where the dielectric function is piecewise constant, 
the group velocity in the presence of material dispersion effects, may be
written\cite{matdisp}:

\vspace{12pt}
\be
v_g=\frac{v_g^0}{1+\frac{\om}{2}E_d\dloge}, 
\label{vgeq}
\ee
\vspace{12pt}

where $E_d$ is the fraction of the electric-field energy present in the 
dielectric and $v_g^0$ is the group velocity in the absence of material 
dispersion. The latter may be calculated directly from the fields as\cite{snyderandlove}:

\vspace{12pt}
\be
v_g^0=c\frac{{\mathrm Re}\langle \left[{\bf E^{\ast}\times H}\right]_z\rangle}{\langle{\bf H,H}\rangle}
\label{vg0}
\ee
\vspace{12pt}

Thus, the group velocity $v_g$ can be evaluated directly from the fields once the
guided mode has been obtained, and the dispersion coefficient can then be
calculated by a numerical first-order derivative. This procedure requires
that $\om, E_d$ and $v_g^0$ are evaluated at the silica refractive index
appropriate for $\om$, which in the present work is achieved by a 
self-consistency procedure\cite{barkou-pbg-disp}. The self-consistent calculations are compared
with calculations assuming a fixed value of the silica refractive index, 
$n$, in order to assess the importance of material dispersion effects.
In the selfconsistent calculations we use the Sellmeier formula for the 
frequency dependence of the silica refractive index, with the coefficients
reported by Okamoto\cite{okamoto}. 

\section{Numerical results}
In this work we focus on the guidance of the fundamental mode (whose major
transverse part is circularly symmetric) in the lowest bandgap. Initially,
we will consider the case of a fixed silica refractive index $n$=1.45. For this
value of $n$, the fiber with
\dbl=0.88 is found to have a narrow transmission window for the fundamental 
mode between \lbl=0.724 and \lbl=0.685, whereas the fiber with \dbl=0.95 has 
a somewhat wider transmission window between \lbl=0.617 and \lbl=0.533.
 As will become clear later these
transmission windows show some dependence on the material refractive index,
which translates into a dependence on the physical value of the pitch (since
this controls the physical wavelength of the light in the guided mode).
In both fiber designs higher-order modes are present in part of the transmission
range of the fundamental mode. For \dbl=0.88 we find that second-order modes are present
in the fundamental bandgap in the lower three-quarters of the transmission window for
the fundamental mode. For \dbl=0.95 the second-order modes leave the bandgap
somewhat earlier, when the fundamental mode is roughly in the middle of the
bandgap. Since the question of determining the single-mode wavelength regions of
the fibers is complicated by the possibility of guidance in the
higher-order bandgaps, and is not a primary concern in this paper, we have not 
attempted a precise determination of the transmission windows for the 
second-order modes. 

\begin{figure}[h]
\resizebox{10cm}{!}{\includegraphics[0cm,0cm][20cm,29cm]{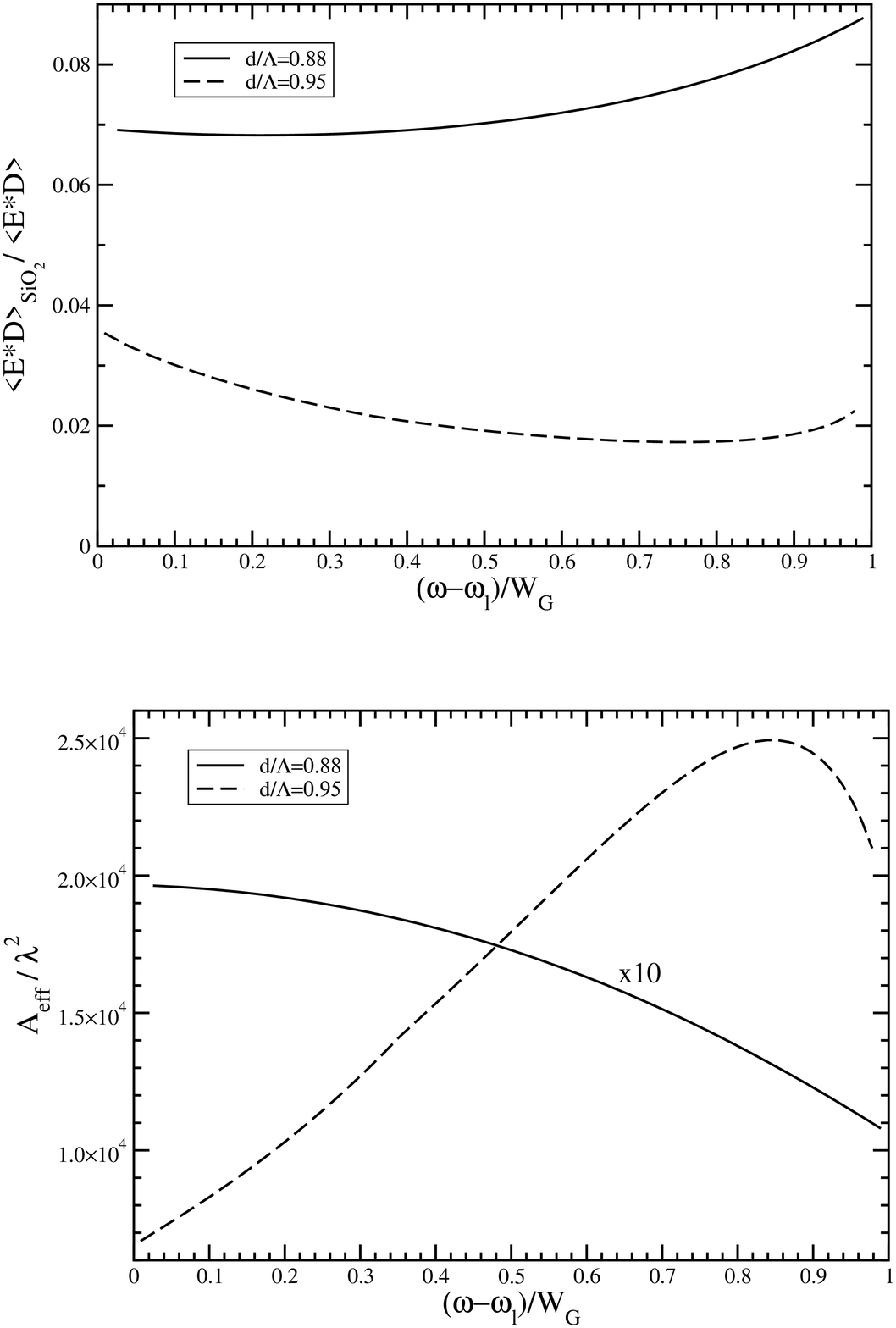}}
\caption{}
\label{figaeff}
\end{figure}

  In Fig.~\ref{figaeff}(a) the fraction of the electric field energy present 
in the silica part of the fibers ($E_d$ in Eq. (\ref{vgeq})) is plotted 
as a function of the distance between the frequency of the fundamental mode 
and the lower band-gap edge normalized to the gapwidth. Both fiber designs 
show the same qualitative behaviour: $E_d$ rises as the mode enters or
leaves the gap, and therefore a minimum is present inside the transmission
window. However, for the design with \dbl=0.88 the minimum is present in 
the low-frequency part of the transmission window, whereas for \dbl=0.95 the
minimum is shifted close to the high-frequency transmission edge. It is
also noteworthy that the frequency derivative of $E_d$ 
is quite large, since the transmission windows are narrow. This has important
consequences for the dispersion properties of the fibers. 

  In Fig.~\ref{figaeff}(b) the effective areas, as calculated from Eq. (\ref{aeff}), are
plotted for the two fiber designs. The results for \dbl=0.88 have been
multiplied by a factor of 10 to facilitate comparison. As expected, very
large \aeff values are found, signifying very low nonlinear coefficients. 
In index guiding microstructured fibers in the large-mode area regime 
($\lambda \ll \Lambda$) 
the effective area $A_{\rm eff} \sim \alpha\times (d/\Lambda)^{-1}\Lambda^2$ 
with a numerical prefactor $\alpha$ of the order 0.5 \cite{mortensen2002a}. 
The fibers are typically operated close to the endlessly-single mode limit 
($d/\Lambda\sim 0.45$) so that $A_{\rm eff}\sim \Lambda^2$. Typical values
of $\Lambda$ are 10-20 $\mu$m, so that $A_{\rm eff}\sim$100$\lambda^2$ for
$\lambda\sim$1$\mu$m. Thus, the present results for airguiding PBG fibers indicate 
a lowering by 1-2 orders of magnitude of the nonlinear coefficient compared 
to typical index-guiding large-mode area microstructured fibers available.
Still, it is interesting to observe the significant variation of \aeff
over the transmission window, and the strong dependence on cladding
design of the nonlinear coefficients. In the fiber with \dbl=0.88 a decrease
of the effective area with increasing frequency is seen, corresponding to
the increasing fraction of field energy in silica (see Eq. (\ref{aeff})). 
For \dbl=0.95 the opposite trend occurs, due to the shift of the minimum
in $E_d$. Of course, the effective areas reported here
relate to the nonlinearity coefficients and have little to do with the
physical size of the modes. This is better estimated from the
standard definition of effective area, Eq. (\ref{oldaeff}), which for both
fiber designs is found to be comparable to the area of the hollow core,
indicating that the guided mode is well localized. 

\begin{figure}[h]
\resizebox{13cm}{!}{\includegraphics[0cm,0cm][20cm,29cm]{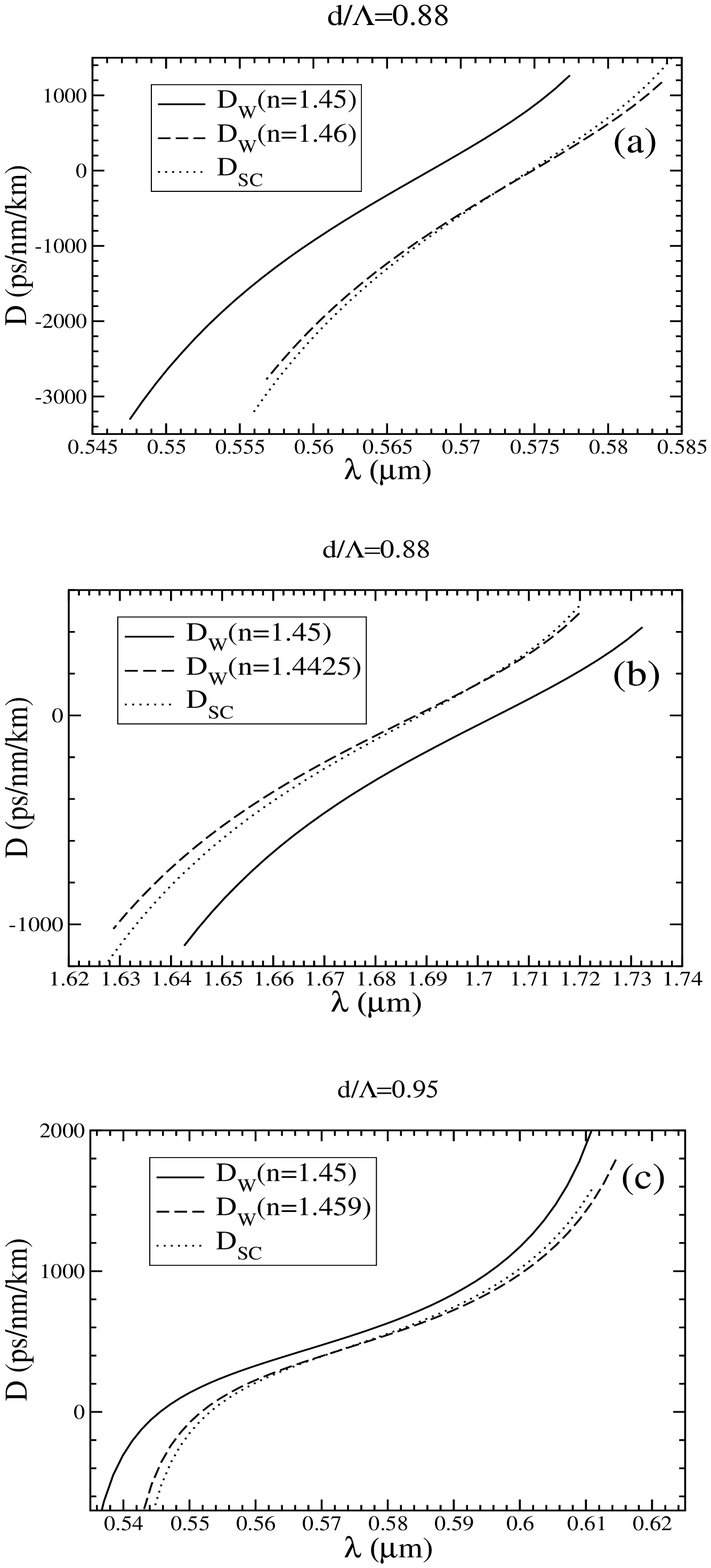}}
\caption{}
\label{figgvd}
\end{figure}

 In Fig. ~\ref{figgvd} GVD results for  three different fiber designs are 
reported. For the fiber with $d/\Lambda$=0.88 we have investigated two values
of the pitch, $\Lambda$=0.8 $\mu$m and $\Lambda$=2.4 $\mu$m. For the design
with $d/\Lambda$=0.95 we show results for $\Lambda$=1.0 $\mu$m. Both the
results of self-consistent calculations and of calculations with a fixed
value of the silica dielectric constant are shown. It can be seen that a
change in the silica refractive index, $n$, shifts the transmission windows, 
and thereby the dispersion curves. Due to the steepness of the dispersion 
curves this implies that the dispersion at a given wavelength is strongly 
dependent on $n$. Therefore, the waveguide GVD calculated at $n$=1.45 (the
solid curves), which is the refractive index of silica at a wavelength of
1.05 $\mu$m, gives a poor prediction of the true chromatic dispersion (as
given by the self-consistent calculations, reported by the dotted curves)
at other wavelengths. The agreement is considerably improved by choosing
a fixed index suitable for the wavelength of the guided mode. The results
of such calculations are reported by the dashed curves. However, there is
still a noticeable difference between the dispersion curves calculated at
a fixed $n$ and the self-consistent results. The differences are of the
same order of magnitude as the material dispersion of homogeneous silica at
the wavelengths in question, and are seen to change sign over the transmission
window. In Fig.~\ref{figgvd}(a) it is interesting to notice that the dashed
and dotted curves do not tend towards each other at the shortest wavelengths
of guidance even though the value of $n$=1.46 used for calculation of the 
dashed curve corresponds to a wavelength of $\sim$550 nm for pure silica.
Instead, the curves cross at somewhat longer wavelength. These findings 
indicate that material dispersion effects play a significant role despite the
small percentage of field energy present in silica.

  In order to obtain a more detailed understanding of the influence of 
material dispersion, we return to Equations (\ref{deq}) and (\ref{vgeq}).
The derivative of the group velocity with respect to frequency may be
written:

\vspace{12pt}
\be
\frac{d v_g}{d \om}=\frac{v_g}{v_g^0}\left(\frac{1}{v_g^0}\frac{\partial
v_g^0}{\partial\beta}+\frac{\partial v_g^0}{\partial\eps}\frac{d\eps}
{d\om}-v_g\left(\frac{E_d}{2}\frac{d\ln\eps}{d\om}+\frac{\om}{2}
\left(\frac{d E_d}{d\om}\frac{d\ln\eps}{d\om}+E_d\frac{d^2\ln\eps}{d\om^2}
\right)\right)\right)
\ee
\vspace{12pt}

 In this formula, $\partial/\partial\om$ ($\partial/\partial\beta$) denotes a derivative with respect
to $\om$ ($\beta$) for a fixed value of $\eps$, whereas $d/d\om$ ($d/d\beta$) 
denotes a derivative including the variation of $\eps$ with $\om$ (and 
thereby $\beta$).  If dispersion in the base material of the fiber is
neglected only the first term contributes. Using the equation\cite{matdisp}:

\vspace{12pt}
\be
\frac{\partial v_g^0}{\partial\eps}= -\frac{E_d}{2\eps}
v_g^0-\frac{\om}{2\eps}\frac{\partial E_d}{\partial\beta}
\ee
\vspace{12pt}

and approximating $dE_d/d\om\approx\partial E_d/\partial\om$, which we have
found to be reasonably well justified even for the airguiding fibers studied
here, we can write:

\vspace{12pt}
\bea
\frac{d v_g}{d \om}&=&\frac{v_g}{\left(v_g^0\right)^2}\frac{\partial
v_g^0}{\partial\beta}-\frac{(v_g)^2}{v_g^0}E_d\left
(\dloge\left(1+\frac{\om}{4}E_d\dloge\right)+\frac{\om}{2}\dtloge\right)-\nonumber\\\nonumber\\
& &\frac{(v_g)^2}{v_g^0}\om\dpedw\dloge\left(1+\frac{\om}{4}E_d\dloge\right)
\label{dvgdweq}
\eea
\vspace{12pt}

  Thus the GVD may be separated into a part independent of material dispersion
effects (first term in Eq. (\ref{dvgdweq})), a part proportional to $E_d$
and a part proportional to the frequency derivative of $E_d$. Using Eq. 
(\ref{deq}) the GVD is found to be: 

\vspace{12pt}
\bea
D&=& D^{SC}_w - \frac{\om^2E_d}{2\pi cv_g^0}\left(\dloge\left(1+\frac{\om}{4}E_d\dloge\right)+\frac{\om}{2}\dtloge \right)\nonumber\\\nonumber\\
&&-\frac{\om^3}{2\pi cv_g^0}\dpedw\dloge\left(1+\frac{\om}{4}E_d\dloge\right)
\equiv D_w^{SC}+D_{mat}
\label{matdispeq}
\eea
\vspace{12pt}

where $D^{SC}_w$ is the GVD in the absence of material dispersion, but 
evaluated at the silica refractive index appropriate for the $\om$ value
in question. In Fig.~\ref{figmatdisp}
the material dispersion $D_{mat}$ defined by Eq. (\ref{matdispeq}) is plotted
for the two designs guiding at short wavelengths. For the design with
$d/\Lambda$=0.88, where 7-9\% of the field energy is in silica, the material
dispersion ranges between $\sim$0 and -200 ps/nm/km, whereas for the design
with $d/\Lambda$=0.95, and only 2-3\% of the field energy in silica, the
material contribution to the GVD ranges between -50 and 50 ps/nm/km. Also 
shown in Fig.~\ref{figmatdisp} is the difference between the waveguide 
dispersion at fixed, suitably chosen, $n$ (dashed lines in Fig.~\ref{figgvd})
and the self-consistent dispersion coefficients (dotted lines in
Fig.~\ref{figgvd}). Exact correspondence between solid and dashed curves in
Fig.~\ref{figmatdisp} is not to be expected since the waveguide dispersion,
$D^{SC}_w$ in Eq. (\ref{matdispeq}) is evaluated at the self-consistent value
of $n$, however, it can be seen that the major part of the discrepancy between
$D_W$ and $D_{SC}$ in Fig.~\ref{figgvd} can be attributed to the intrinsic
material dispersion as expressed by $D_{mat}$.
Since the material dispersion of homogeneous silica in this wavelength range is between -250 and -400 ps/nm/km, 
the $D_{mat}$ values reported in Fig.~\ref{figmatdisp} are surprisingly large 
considering the small values of $E_d$.

It is evident from Eq.  (\ref{matdispeq}) that the contribution of material effects to the total
GVD of a fiber is composed of a part proportional to $E_d$ and a part
proportional to $\partial E_d/\partial\om$. Herein lies the origin of the
surprisingly large $D_{mat}$ values for the airguiding
fibers: Although $E_d$ is small in these fibers, as is evident from
Fig.~\ref{figaeff}(a), this is not the case for $\partial E_d/\partial\om$.
In Fig.~\ref{figrd} we plot the ratio, $R_D$ between the third and second
term in Eq. (\ref{matdispeq}):

\vspace{12pt}
\be
R_D=\frac{\om\dplogeddw\dloge}{\dloge + \frac{\frac{\om}{2}\dtloge}{1+\frac{\om}{4}E_d\dloge}}
\label{rdeq}
\ee
\vspace{12pt}

The results for the airguiding fibers in Fig.~\ref{figrd}(a) are compared
to results for index guiding microstructured fibers, reported in 
Fig.~\ref{figrd}(b). The latter have a triangular cladding structure
similar to the airguiding fibers, but with a solid silica core defined by
a missing airhole. It can be seen that $\mid R_D\mid$ for the index guiding fibers
is everywhere below unity, even in the rather extreme case of $d/\Lambda$=0.8,
$\Lambda$=0.34 $\mu$m, where $\sim$15\% of the field energy is located in
the airholes. For the airguiding PBG fibers $\mid R_D\mid$ is 1-2 orders of 
magnitude larger. Thus, for index-guiding microstructured
fibers the main contribution to material dispersion effects comes from the
second term in Eq. (\ref{matdispeq}), whereas for airguiding PBG fibers
the contribution from the third term dominates. 

  Because airguiding PBG fibers have the major part of the field energy
propagating in air, the choice of base material is less limited by requirements
of low loss and/or nonlinearity than is the case for standard fibers, or
index-guiding microstructured fibers. On the other hand, the results presented
in this work show that the dispersion properties of the material may
still have a significant impact on the total GVD of the fiber. These 
observations suggest that dispersion engineering through the choice of base
material may be a possibility in these fibers. A simple example of the
possibilities is shown in Fig.~\ref{figdoped}: The usual three-term Sellmeier 
polynomial describing the material dispersion of silica has been modified
by a fourth term describing the addition of an (dopant) absorption line close
to the transmission window of the fiber. The modified Sellmeier polynomial
reads:

\vspace{12pt}
\bea
\eps (\lambda ) &=& 1+\sum_{i=1}^4 \frac{a_i}{\lambda^2-\lambda_i^2}
\\\nonumber\\
a_1=0.6965325 \mu m^{-2} && \lambda_1=0.066 \mu m\nonumber\\\nonumber\\
a_2=0.4083099 \mu m^{-2} && \lambda_2=0.118 \mu m\nonumber\\\nonumber\\
a_3=0.8968766 \mu m^{-2} && \lambda_3=9.896 \mu m\nonumber\\\nonumber\\
a_4=0.001 \mu m^{-2} &&\nonumber 
\eea
\vspace{12pt}

with $\lambda_4$ chosen as either 0.5 $\mu$m (P500 in Fig.~\ref{figdoped})
or 0.640 $\mu$m (P640 in Fig.~\ref{figdoped}). It is evident that significant
shifts of the dispersion curve can be obtained simply by addition of (impurity)
absorption centers to the silica matrix. A more general approach would of
course be to vary the composition of the base material itself as could readily
be done in, e.g., polymer fibers. Such dispersion engineering could, for 
instance, be of interest for fibers applied to the kind of gas-phase 
nonlinearity experiments whose feasibility was recently demonstrated by 
Benabid and co-workers\cite{benabid2002b}.

\section{Conclusions}

  In conclusion, we have investigated various aspects of the interplay 
between base material and the fundamental guided mode in silica-based
airguiding PBG fibers. For the two designs studied here, between 2 and
9\% of the electric field energy was found to reside in the silica parts
of the fiber. The nonlinearity coefficient was expressed in terms of a 
generalized effective area, which was found to be 1-2 orders of magnitude
larger than what can be obtained in index-guiding microstructured fibers.
The influence of material dispersion on the total GVD of the fibers was
investigated and was found to be of the same order of magnitude as in
other fiber types having most of the field energy residing in silica. This
effect was traced to the fact that the variation with frequency of the field 
energy in silica is much more rapid in airguiding PBG fibers than in other
fiber types. These results suggest that dispersion engineering through
the choice of base material may be an interesting possibility in airguiding 
PBG fibers.


\begin{thebibliography}{10}

\bibitem{cregan1999}
R.~F. Cregan, B.~J. Mangan, J.~C. Knight, T.~A. Birks, P.~St.~J. Russell, P.~J.
  Roberts, and D.~C. Allan.
\newblock Single-mode photonic band gap guidance of light in air.
\newblock {\em Science}, 285:1537--1539, 1999.

\bibitem{benabid2002a}
F.~Benabid and J.~C. Knight P. St.~J. Russell.
\newblock Particle levitation and guidance in hollow-core photonic crystal
  fiber.
\newblock {\em Optics Express}, 10:1195--1203, 2002.

\bibitem{ouyang2002}
G.~Ouyang, Y.~Xu, and A.~Yariv.
\newblock Theoretical study on dispersion compensation in air-core bragg
  fibers.
\newblock {\em Optics Express}, 10:899--908, 2002.

\bibitem{benabid2002b}
F.~Benabid, J.~C. Knight, G.~Antonopoulos, and P.~St.~J. Russell.
\newblock Stimulated raman scattering in hydrogen-filled hollow-core photonic
  crystal fiber.
\newblock {\em Science}, 298:399--402, 2003.

\bibitem{venkataraman2002}
N.~Venkataraman, M.~T. Gallagher, C.~M. Smith, D.~M{\"u}ller, J.~A. West, K.~W.
  Koch, and J.~C. Fajardo.
\newblock Low loss (13 db/km) air core photonic bandgap fiber, {28th European
  Conference on Optical Communication, ECOC '02, September 2002, Copenhagen,
  Denmark, post-deadline paper PD1.1}.

\bibitem{broeng-ag}
J.~Broeng, S.~E. Barkou, T.~S{\o}ndergaard, and A.~Bjarklev.
\newblock Analysis of air-guiding photonic bandgap fibers.
\newblock {\em Opt. Lett.}, 25:96--8, 2000.

\bibitem{white-ag-mp}
T.~P. White, R.~C. McPhedran, L.~C. Botten, G.~H. Smith, and C.~Martijn
  de~Sterke.
\newblock Calculations of air-guided modes in photonic crystal fibers using the
  multipole method.
\newblock {\em Optics Express}, 9:721--32, 2001.

\bibitem{mpb}
S.~G. Johnson and J.~D. Joannopoulos.
\newblock Block-iterative frequency-domain methods for {M}axwell's equations in
  a planewave basis,.
\newblock {\em Optics Express}, 8:173--190, 2001.

\bibitem{matdisp}
J.~L{\ae}gsgaard, A.~Bjarklev, and S.~E.~Barkou Libori.
\newblock Chromatic dispersion in photonic crystal fibers: Fast and accurate
  scheme for calculation.
\newblock {\em J. Opt. Soc. Am. B}, 20: 443, 2003.

\bibitem{agrawal}
G.~P. Agrawal.
\newblock {\em Nonlinear Fiber Optics}.
\newblock Academic Press, San Diego, 2001.

\bibitem{pbg_ea}
J.~L{\ae}gsgaard, N.~A. Mortensen, and A.~Bjarklev.
\newblock Mode areas and field energy distribution in honeycomb photonic
  bandgap fibers, accepted for {J. Opt. Soc. Am. B} [physics/0307078]

\bibitem{snyderandlove}
A.~W. Snyder and J.~D. Love.
\newblock {\em Optical Waveguide Theory}.
\newblock Chapman \& Hall, London, 1996.

\bibitem{barkou-pbg-disp}
S.~E. Barkou, J.~Broeng, and A.~Bjarklev.
\newblock 'dispersion properties of photonic bandgap guiding fibers,' {Optical
  Fiber Communication Conference, pp. 117-9, paper FG5, San Diego, Feb. 1999}.

\bibitem{okamoto}
K.~Okamoto.
\newblock {\em Fundamentals of optical waveguides}.
\newblock Academic Press, San Diego, 2000.

\bibitem{mortensen2002a}
N.~A. Mortensen.
\newblock Effective area of photonic crystal fibers.
\newblock {\em Opt. Express}, 10:341--348, 2002.

\end{thebibliography}

\newpage

\section*{List of Figure Captions}

Fig. 1. Schematic picture of one of the structures (with $d/\Lambda$=0.88)
under study. The black circles are airholes, while the white areas are the
silica regions. Only the core and innermost cladding region is shown.

Fig. 2. Field energy fraction in silica (a) and effective area as calculated
from Eq. (\ref{aeff}) (b). The effective area curve for the fiber with 
$d/\Lambda$=0.88 (solid curve in (b)) has been multiplied by 10 to facilitate comparison.

Fig. 3. Dispersion curves for the fundamental guided mode of three airguiding
PBG fibers with various values of $d$and $\Lambda$. (a): $d/\Lambda$=0.88,
$\Lambda$=0.8 $\mu$m. (b): $d/\Lambda$=0.88, $\Lambda$=2.4 $\mu$m. 
(c): $d/\Lambda$=0.95, $\Lambda$=1.0 $\mu$m. Solid curves report waveguide 
dispersion (D$_W$) calculated at $n$=1.45, dashed curves report waveguide
dispersion at values of $n$ suitable for the wavelength interval spanned by
the transmission window, and dotted curves denote results of self-consistent
calculations.

Fig. 4. Material dispersion $D_{mat}$ (solid lines), defined in Eq. (\ref{matdispeq}), for 
the two fiber designs guiding at short wavelengths. The dashed lines report the
difference between the dotted and dashed curves in Fig. 3 for comparison.

Fig. 5. Plots of the quantity $R_D$, defined in Eq. (\ref{rdeq}), for 
two airguiding PBG fiber designs (a) and two index-guiding fibers (b) having
a cladding structure similar (although with smaller airholes) to the
airguiding PBG fibers.

Fig. 6. Dispersion curves for two fibers with added absorption resonances
in the base material at either $\lambda$=500 nm (P500) or $\lambda$=640 nm 
(P640) compared to the undoped result.

\newpage

\begin{figure}[h]
\resizebox{13cm}{!}{\includegraphics[0cm,0cm][20cm,29cm]{fig4.eps}}
\caption{}
\label{figmatdisp}
\end{figure}

\begin{figure}[h]
\resizebox{12cm}{!}{\includegraphics[0cm,0cm][20cm,29cm]{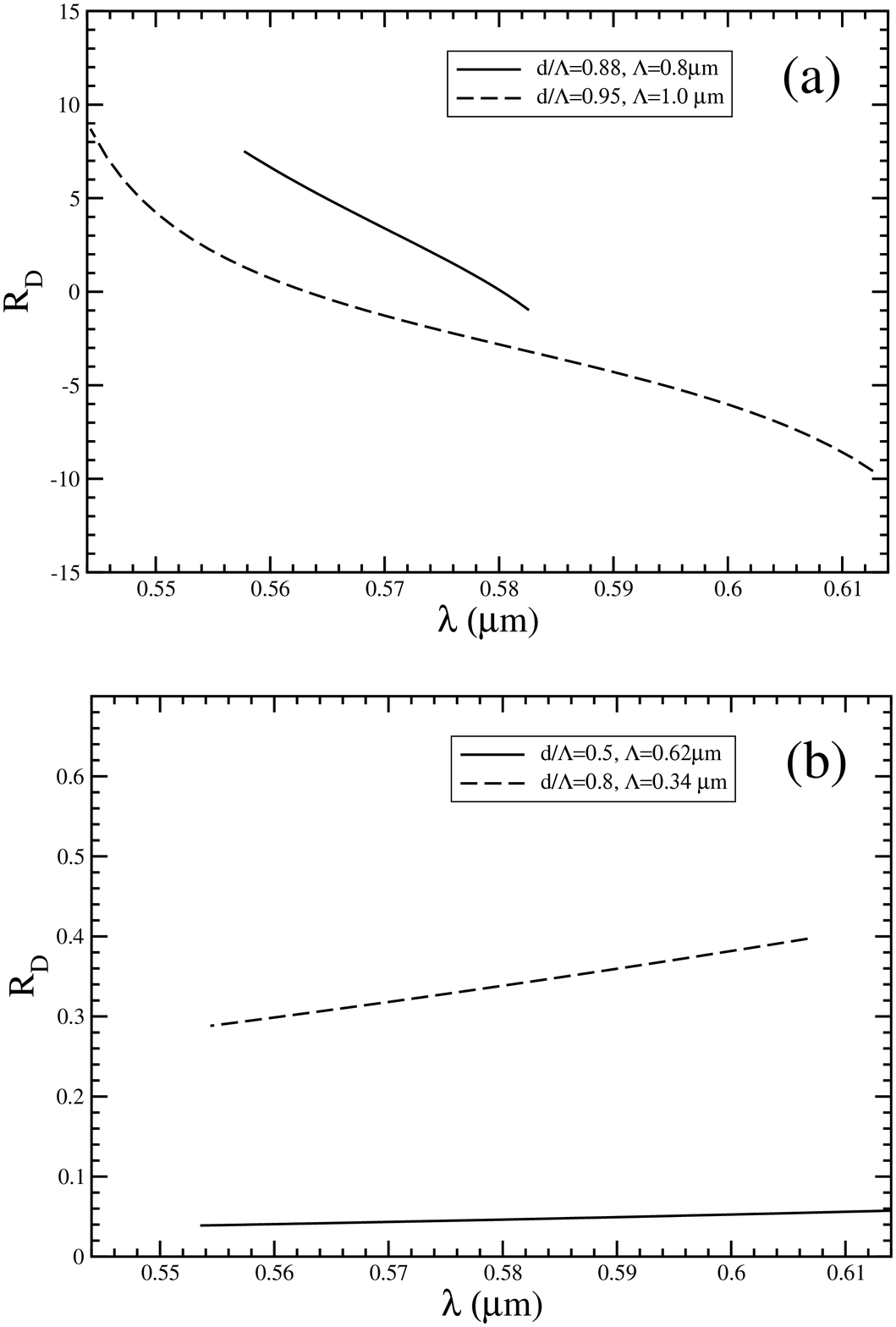}}
\caption{}
\label{figrd}
\end{figure}

\begin{figure}[h]
\resizebox{12cm}{!}{\includegraphics[0cm,0cm][20cm,29cm]{fig6.eps}}
\caption{}
\label{figdoped}
\end{figure}

\end{document}